\documentclass{emulateapj}
\shorttitle{Radiative Thrusters}
\shortauthors{FABRYCKY}
\slugcomment{Accepted to ApJ Letters: March 12, 2008}

\begin{document}
\title{Radiative Thrusters on Close-in Extrasolar Planets}
\author{Daniel Fabrycky\altaffilmark{1}}
\affil{Harvard-Smithsonian Center for Astrophysics\\
	60 Garden St, MS-51, Cambridge, MA 02138}
\email{daniel.fabrycky@gmail.com}
\altaffiltext{1}{Michelson Fellow}

\begin{abstract}
The atmospheres of close-in extrasolar planets absorb most of the incident stellar radiation, advect this energy, then reradiate photons in preferential directions.  Those photons carry away momentum, applying a force on the planet.  Here we evaluate the resulting secular changes to the orbit, known as the Yarkovsky effect.  For known transiting planets, typical fractional changes in semi-major axis are about 1\% over their lifetime, but could be up to $\sim 5\%$ for close-in planets like OGLE-TR-56b or inflated planets like TrES-4.  We discuss the origin of the correlation between semi-major axis and surface gravity of transiting planets in terms of various physical processes, finding that radiative thrusters are too weak by about a factor of 10 to establish the lower boundary that causes the correlation.  
\end{abstract}

\section{Introduction}

The radiative environment of a close-in extrasolar planet is intense.  At the canonical hot-Jupiter distance of $0.05$~AU around a solar-type star, the incoming flux is a factor of $10^4$ larger than that experienced by Jupiter itself.  The starlight incident on a hot Jupiter is mostly absorbed, as very strong upper limits on reflected starlight attest: the geometric albedo of HD209458b at visible wavelengths is less than 0.17 at 99.5\% confidence \citep{2007R}, much smaller than Jupiter's value of 0.5.  As the bulk of the planet is expected to be tidally locked, the time-averaged radiation field is strongly anisotropic.  This heat input drives winds around the planet that are of planetary scale.  

The atmospheric dynamics on hot Jupiters is a hot topic, for both theory and observation.  A consistent theoretical prediction of three-dimensional models is that, in a steady state, the hottest portion of the planet does not lie at the substellar point, but is shifted eastward by a superrotating wind \citep{2002ShowmanG}.  The theoretical interpretation of this feature of the simulations is still uncertain, but \cite{2007SMC} suggest the day-night temperature contrast triggers eddies that drive eastward angular momentum to the equator.  Observations are starting to reveal the existence and extent of the phase shift.  A hotspot displaced to the east by $\sim30^\circ$ was recently observed at $8~\mu$m on the planet HD189733b by \cite{2007K}.  \citet{2006H} observed the phase curve of $\upsilon$~And~b to have a best-fit westward shift of $\sim11^\circ$, but the data are roughly consistent with no shift.  

The orbital mechanics of hot Jupiters depend on the radiation from their atmospheres.  Because light carries momentum, it exerts a force as it is absorbed or emitted anisotropically.  Absorbing the incident light pushes the planet away from the star.  Since the amount of radiation intercepted is proportional to the inverse square of separation, the effect is the same as a smaller stellar mass, but the orbit remains Keplerian.  On the other hand, if the radiation from the planet is preferentially tangential to the orbit, its semi-major axis can increase or decrease.  If it is true that the hotspot moves eastward on average, the reradiated light acts as a thruster back along the path of the planet, adding angular momentum to the planet, which promotes it to a higher orbit. 

Thermal radiation acting as a thruster which changes the orbit is well-known in asteroid dynamics as the Yarkovsky effect \citep{1951O, 2006B}.  In that case, sunlight heats up the day side, but the asteroid's spin is not synchronous with its orbit, and rotation displaces the energy before it is reemitted.  This effect causes substantial orbital changes over the age of the solar system only for small ($<10$~km) asteroids due to their large cross-section per unit mass.  Relative to a 10 km radius asteroid in the main belt of the solar system, hot Jupiters are $\sim 10^{12}$ times more massive.  However, they are also $\sim 10^2$ times closer to the star and have a radius $\sim 10^4$ times larger; i.e., they intercept $10^{12}$ times the luminosity.  Therefore, if the reradiation occurs with similar phase angle and contrast, the fractional orbital changes of planets close to other stars will be the same as that of asteroids of the solar system.

The plan of this \emph{Letter} is as follows.  In \S\ref{sec:efforbit} we show the rate of change of orbital parameters according to celestial mechanics.  In \S\ref{sec:calc} we calculate these effects for two model planetary atmospheres and an observed planet.  In \S\ref{sec:correl} we compare the observed distribution of transiting planets to limits due to various physical processes that tend to move or destroy them.  Finally, \S\ref{sec:discuss} closes with a brief discussion and call for further study.

\section{Effect on the orbit} \label{sec:efforbit}
	Here we compile formulae that give the effects of a force on an orbit, following \cite{1999MD}, section 2.9.  The force imparted by the radiation can be decomposed into radial, transverse, and normal components: $\mathbf{F} = R \mathbf{\hat{r}} + T \mathbf{\hat{\theta}} + N \mathbf{\hat{z}}$.  
	
	The rate of change of semi-major axis is:
		\begin{equation}
		\dot{a} = \frac{2 a^{3/2}}{m_p \sqrt{G (M_\star+m_p) (1-e^2)}} [R e \sin f + T (1 + e \cos f)], \label{eq:dadt_md}
		\end{equation}  
where $f$ is the true anomaly, $e$ is eccentricity, and $M_\star$ and $m_p$ are the stellar and planetary masses.  The thrust is produced by the dipole of the reradiated flux distribution, which we characterize by its fractional amplitude $f_d$, so that the flux distribution has an angular dependence $\propto (1+f_d \cos \psi)$, where $\psi$ is measured from a pole through the peak of the dipole.  The incoming starlight is Doppler-boosted in the frame of the planet, and upon absorption causes angular momentum loss (Poynting-Robertson drag), but that effect is $v/c \sim 5\times10^{-4}$ weaker, where $v$ is the orbital velocity and $c$ is the speed of light.  The force due to the dipole component is 
		\begin{equation}
		F = \frac{\mathcal{L}_\star (1-A)}{8 c} \Big(\frac{R_p}{r}\Big)^2 f_d ,
		\end{equation}
where $\mathcal{L}_\star$ is the star's luminosity, $R_p$ is the planet's radius as measured by photometry during transit, $r = a(1-e^2)(1+e \cos f)^{-1}$ is the distance between the planet and star, and $A$ is the planet's albedo.  (We assume that the albedo does not vary over the surface of the planet; if it does, there is additional opportunity for momentum transfer between starlight and the orbit.)  We assume atmospheric dynamics displace the dipole a longitudinal angle $\lambda_d$ east of the substellar point, so the components of this force are: $R = F \cos \lambda_d$, $T = F \sin \lambda_d$, and $N = 0$.  For an order-of-magnitude estimate, let us assume $\lambda_d$ and $f_d$ are constant. There may be an anti-correlation in the magnitudes of $\lambda_d$ and $f_d$ as high-altitude opacity sources may cause planets with high effective temperature to reradiate the energy before it is advected \citep{2007Fb}.  Averaging equation~(\ref{eq:dadt_md}) over time, we have:
		\begin{equation}
		\dot{a} = \frac{\mathcal{L}_\star (1-A) f_d \sin \lambda_d R_p^{2} }{4 c  G^{1/2} (M_\star+m_p)^{1/2} m_p a^{1/2} (1-e^2) }. \label{eq:dadt}
		\end{equation}  
Therefore the timescale for semi-major axis change is:
		\begin{eqnarray}
		\tau_a &\equiv& a / \dot{a} \nonumber \\
		&=& 5.5 \times 10^{11}~{\rm yr} \Big( \frac{f_d \sin \lambda_d}{0.5} \Big)^{-1}  \Big(\frac{\mathcal{L}_\star (1-A) }{\mathcal{L}_\odot}\Big)^{-1}  \Big(\frac{m_p R_p^{-2}}{m_{J} R_J^{-2}}\Big) \nonumber \\
		& &\qquad \times (1-e^2) \Big(\frac{a}{0.05 {\rm AU}}\Big)^{3/2}  \Big(\frac{M_\star+m_p}{M_\odot} \Big)^{1/2},   \label{eq:taua} 
		\end{eqnarray}
where reference values are for Jupiter and the Sun.  Therefore the timescale of orbital change for a typical hot Jupiter is about $100$ times its age.

	The rate of change in eccentrity is: 
		\begin{equation}
		\dot{e} = \frac{1}{m_p}\sqrt{\frac{a (1-e^2)}{G (M_\star+m_p)}} [R \sin f + T (\cos f + \cos \mathcal{E})],\label{eq:dedt}
		\end{equation}
where $\mathcal{E}$ is the eccentric anomaly.  Now, using the same force and assumptions as above, we have:
		\begin{equation}
\tau_e \equiv (d\ln e/dt)^{-1} = \tau_a \frac{2 e^2}{1-e^2-(1-e^2)^{3/2}} \approx 4 \tau_a,
 		\end{equation}
where the final approximation is first-order in eccentricity.  The Yarkovsky effect causes eccentricity to grow if the hotspot is shifted to the east (e.g., \citealt{2002SG}), on a timescale similar to that of semi-major axis growth.  The assumptions of constant $f_d$ and $\lambda_d$ will break down at large eccentricity, which will be shown in \S\ref{sec:calc}.  However, given that the Yarkovsky effect will only be large for small values of $R_p/a$, tidal dissipation in the planet should be sufficient to damp the eccentricity faster than the Yarkovsky effect excites it.
		
		The rate of change in orbital inclination is: $\dot I \propto N$.  Above we assumed that the reradiation field is anisotropic, but is still symmetric above and below the plane of the orbit.  In reality, the normal component of the force, $N$, fluctuates with fluctuating eddies.  However, this is a random walk with extremely small step size, so it is unlikely to get very far, even if the normalization of the rate were large.  
	
\section{Calculation of the force} \label{sec:calc}
	In this section the rates of orbital evolution are computed for three planets, taking the flux distribution for HD 209458b and HAT-P-2b from atmospheric models and for HD 189733b from infrared observations.

Atmospheric models readily give the effective temperature distribution $T_{\rm{eff}}$ of the planet, each patch of which radiates into a hemisphere.  The thrust from a patch is ${\bf \delta F} = -f' \sigma T_{\rm{eff}}^4 {\bf \delta A} / c$.  Here ${\bf \delta A}$ is the outward normal vector to a patch of area $\delta A$, $\sigma $ is the Stefan-Boltzmann constant, and the factor $f'$ accounts for the fact that the radiation is not all emitted normal to the surface: if it were, $f'$ would be $1$.  If treated as a true blackbody, the radiation from a patch uniformly illuminates the hemisphere, and $f'=\onehalf$.  On the other hand, a limb-darkened atmosphere preferentially emits straight up rather than sideways, so $f'$ is between $\onehalf$ and $1$.  For instance, in a plane-parallel gray atmosphere with intensity proportional to $(\mu+2/3)$, where $\mu$ is the cosine of the angle to the zenith (e.g., \citealt{1978M}), this factor becomes $f'=4/7$.  Integrating ${\bf \delta F}$ over the surface of the planet gives the net force.  Finally,  integrating equations~(\ref{eq:dadt_md}) and (\ref{eq:dedt}) over an orbit gives the secular effect of this force on the orbital elements. 

For the planet HD209458b, \cite{2006CS} carried out a detailed atmospheric circulation model and \cite{2006F} ran radiative transfer calculations to predict phase curves.  For the equilibrium-chemistry case from \cite{2006F}, we compute the dipole component of the flux from the planet (based on the hemisphere-averaged $T_{\rm{eff}}$ as a function of orbital phase) to have an amplitude of $4.9 \times 10^{28}$~erg s$^{-1}$, with its maximum shifted an angle $29^\circ$ eastward.  The timescale for semimajor axis change is then $3.8 \times 10^{11}$~yr; for a fractional change of $\Delta a / a = 0.3\%$ over the age of the system.  (All age estimates are taken from \citealt{2008TWH}.)  The non-equilibrium chemistry case has a darker night side, so the dipole component is bigger, enhancing the Yarkovsky effect by about $20\%$.
	
\begin{figure}
\vspace{0.1 in}
\includegraphics[scale=0.95]{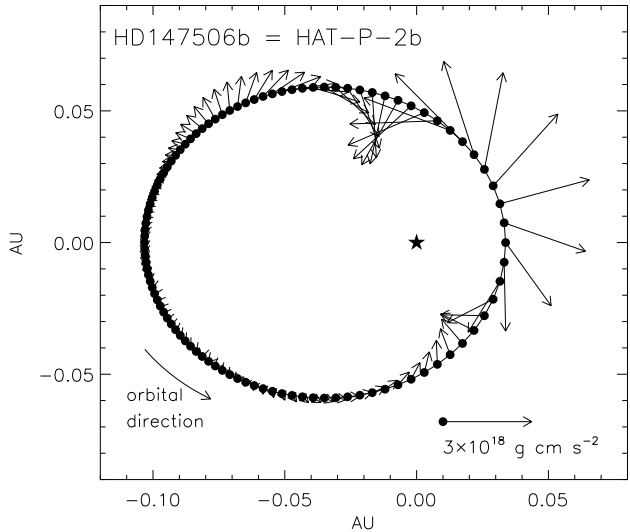}
\caption{ The force vectors due to radiation asymmetry as a function of orbital position for the planet HAT-P-2b, calculated from the atmospheric simulations of \cite{2007LL}.   Only the vector components in the plane of the orbit are shown. \vspace{0.25 in}}
\label{fig:orbforce}
\end{figure}

\cite{2007LL} have computed theoretical atmosphere models for a number of eccentric hot Jupiters.  Here we analyze their computation for the eccentric planet HAT-P-2b, choosing it because it (1) has a rather large eccentricity (0.5) which complements the analysis in \S\ref{sec:calc} and (2) is orbiting a bright star and will therefore receive detailed observations.  See \cite{2007LL} for the details of the model; we additionally assume is that there is no limb darkening.  The model was run for 19.5 orbits to reach dynamical equilibrium, after which we computed the force as a function of time over a single orbit.  Figure~\ref{fig:orbforce} shows that force as a function of orbital position; neither $\lambda_d$ nor $f_d$ are constant.  This force causes growth in both semi-major axis and eccentricity with timescales of $7.6 \times 10^{12}$~yr and $1.1 \times 10^{14}$~yr, respectively, considerably longer than the estimated age of $2.6 \times 10^{9}$~yr.  The rather large eccentricity causes the force to lose coherence over the orbit, and $\tau_e \gg \tau_a$.  

The planet HD189733b has been observed for about 60\% of its orbit, and the $8~\mu$m-brightness of the planet as a function of longitude has been evaluated in 12 longitudinal bins of constant brightness \citep{2007K}.  This $8~\mu$m ``map'' needs to be converted to a bolometric flux ``map'' to compute the Yarkovsky effect.  Most of the flux comes out within a factor of two in wavelength from $8~\mu$m, so we assume these two maps would be roughly the same, although scaled by the bolometric flux ratio of the planet and the star, rather than that ratio at $8~\mu$m.  We assume an albedo of $0.1$ and no limb darkening.  The net force due to reradiation is $1.7\times10^{19}$ g cm s$^{-2}$ in a direction mostly outward from the star, but tipped in the planet's orbital direction by $\lambda_d = 6^\circ$. This shift is smaller than the eastward displacement of the hottest region of the planet ($\sim 30^\circ$) because the coldest region is in the same hemisphere, lessening the component of the force along the orbit.  The timescale for semi-major axis change is $3.0 \times 10^{12}$~yr, so over $5$~Gyr, this thrust produces a 0.17~\% increase in the semi-major axis.  Stellar models suggest the star is at least a few Gyr old.

In the future, both observational and theoretical work on very close planets should compute and report the magnitude of the radiative thruster effect by the methods illustrated here.

%\vspace{0.3 in}
\section{Gravity/Semi-major axis correlation} \label{sec:correl}

\begin{figure}
\includegraphics[scale=0.8]{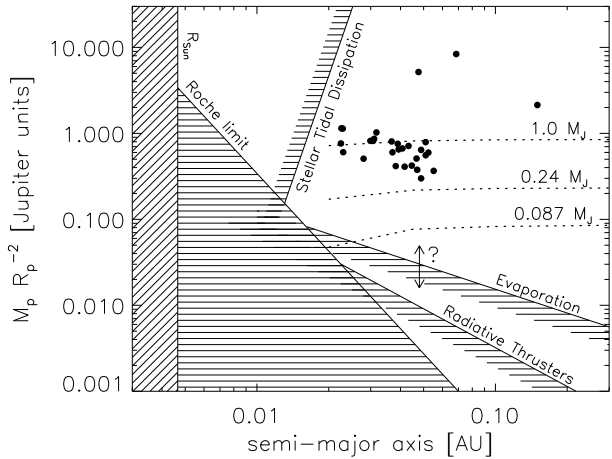}
\caption{ The surface gravity versus semi-major axis for known transiting planets.  Various theoretical mechanisms disfavor the existence of planets on the hashed sides of each labelled line.  The dashed lines are gas-giant models by \cite{2007Fa}; see text. \vspace{0.25 in}}
\label{fig:grav}
\end{figure}

	The scaling of equation~(\ref{eq:taua}) suggests that the relevant property of the planet is its surface gravity.  \cite{2007S} have shown that an anti-correlation exists between surface gravity and semi-major axis for transiting planets (see also \citealt{2005MZP,2008TWH}).  Figure~\ref{fig:grav} is a plot of the relation updated with newly discovered planets.  The sample has a sloping lower boundary, and apart from three outliers, a clear trend.  Lines mark where typical timescales of theoretical effects equal $5$~Gyr, with a hashed area to the ``forbidden'' side of each constraint.	
	
	The Yarkovsky effect, according to equation~(\ref{eq:taua}) using the nominal values of all parameters, is represented by the line labeled ``radiative thrusters''.

	\cite{1996R} raised concerns that very close planets would be pulled \emph{in} towards the star due to tidal dissipation in the star's envelope.  The timescale for tidal decay is \citep{1996R} :
\begin{equation}
\tau_a = 4 \times 10^{13} {\rm ~yr~} f_t \Big(\frac{M_\star}{M_\odot}\Big) \Big(\frac{m_p}{m_J}\Big)^{-1}  \Big( \frac{a / 0.05 {\rm AU}}{R_\star / R_\odot }\Big)^8,
\end{equation}
where $f_t$ has a very uncertain value depending on the tidal dissipation efficiency.  This constraint for a $R_p=R_J$ planet orbiting a solar-type star is plotted in Figure~\ref{fig:grav}.  Although the radiative thruster effect with an eastward-shifted hotspot tends to push the planet out, tides have a much stronger radial dependence, therefore the equilibrium between the two effects is unstable. 
	
	 The tidal gravity of the star limits how close a planet of a given radius can orbit without being disrupted; this ``Roche limit'' is shown in Figure~\ref{fig:grav} for a $1 R_J$ planet on a circular orbit.  \cite{2006FR} have pointed out that the hot Jupiters obey an inner boundary at about twice their Roche limit, which implicates tidal circularization from large eccentricity and semi-major axis.  If that mechanism delivers most hot Jupiters to their current orbits, then it also strongly affects their structure, especially for the least massive ones.  Below a few Jupiter masses, a tidal inflation instability \citep{2003GLB} can result, which can strip the gaseous envelope and push out planets to the Roche limit computed with the severely inflated radius.

	Models of atmospheric escape are very uncertain, but they suggest a hydrodynamic wind is being driven out of the planet's Roche lobe by extreme ultraviolet photons from the star.  \cite{2004Y} found by use of a chemical reaction network that matter can be lost at a rate of $4.7\times10^{10}$~g~s$^{-1}$ for HD209458b.  We may scale this rate to planets of different mass and radius, although several assumptions are necessary.  First, we assume that the optically-thick area of the absorbing species in the exosphere is $A \propto R_p^2$ [probably $A \approx (3 R_p)^2$].  Assuming that the escaping wind uses its energy to climb out of the planet's potential well, $\dot{m}_p \propto A R_p m_p^{-1}$.  Combining with the results for HD209458b gives the following timescale for mass loss:
\begin{eqnarray}
\tau_{m} &\equiv& \frac{m_p}{\dot{m}_p} \nonumber \\
&=& 6 \times 10^{12} {\rm ~yr~} \Big( \frac{m_p}{m_J} \Big)^2 \Big(\frac{R_p}{R_J}\Big)^{-3} \Big(\frac{a }{ 0.05 AU} \Big)^{1.84}.
\end{eqnarray}
Here the power on $a$ is given by Figure~7 of \cite{2004Y}, and is close to 2, the scaling due to a fixed cross section to XUV photons.  This constraint is plotted in the figure as well, for $R_p = R_J$. We note that the stellar XUV luminosity is strongly dependent on the spectral type of the star, which should be taken into account for evaporation \citep{2007L}.  The ``?'' symbol on Figure~\ref{fig:grav} means that the shape and normalization of the plotted limit is still widely debated.
	
	Also plotted in Figure~\ref{fig:grav} are gas giant models of three different masses from Table 4 of \cite{2007Fa}.  The models are of solar metallicity gas with no rocky core, and have been contracting for 4.5 Gyr, while irradiated by a solar-type star (which produces the semi-major axis dependence).  If close-in terrestrial planets can hold on to a thick atmosphere they may have similar atmospheric dynamics as gas giants, along with the Yarkovsky effect.  The models of \cite{2007Fa} for Earth-like compositions and masses $(0.1, 1.0, 10) M_\earth$ have surface gravities of $(0.16, 0.39, 1.22)$ in Jupiter units, so the orbits of terrestrial planets soon detectable by \emph{Kepler} probably have not moved much by this effect.  Such planets without an atmosphere will probably also avoid the asteroid-like Yarkovsky effect (due to spin plus thermal inertia) because of tidal spin synchronization.  
	
\section{Discussion}\label{sec:discuss}
	The goal of this paper is to alert the exoplanet community to a dynamical effect that is very important for understanding the orbits of small asteroids in the solar system and to suggest it may modify the orbits of the closest-in extrasolar planets.
	
	The results of \S\ref{sec:calc} give long timescales for semi-major axis and eccentricity change for known hot Jupiters that have been studied extensively, much longer than their estimated ages, or indeed, the age of the universe.  Of the observed planets, the biggest fractional semi-major axis change is for OGLE-TR-56 and TrES-4, which may have migrated by $\sim 5$~\% during their lifetimes, according to equation~\ref{eq:taua} using the parameters determined by \cite{2008TWH}.  The scalings of the radiative thruster effect imply that it can make a substantial change for the most extreme extrasolar planets, so atmospheric circulation studies should be undertaken for such planets.

	Even a few percent orbital change could enhance the chance of success of current attempts to discover additional planets in mean-motion resonance by transit-timing variations \citep{2005A, 2005HM}.  The orbits of most known transiting planets have likely been modified by tidal dissipation in the planet, which tends to disrupt mean-motion resonance \citep{2007TP}.  However, the radiative thruster effect could oppose this motion, keeping planets in resonance, where transit timing can probe to much smaller masses.

	Finally, we emphasize that the remaining theoretical uncertainties will soon be settled by observations of planetary atmospheres, which will determine the importance of the Yarkovsky effect in the context of extrasolar planets.

\vspace{0.2 in}
I thank J. Langton, H. Knutson, and J. Fortney for supplying me with their data, A. Loeb for numerous discussions, and A. Showman, S. Tremaine, M. Holman, and D. Sasselov for comments.  These ideas were developed during a Michelson Fellowship supported by the National Aeronautics and Space Administration and administered by the Michelson Science Center.

\bibliography{ms} \bibliographystyle{apj}
%\bibliography{yark} \bibliographystyle{apj}

\clearpage 

%\begin{figure}
%\includegraphics{f1.eps}
%\caption{ The force vectors due to radiation asymmetry as a function of orbital position for the planet HAT-P-2b, calculated from the atmospheric simulations of \cite{2007LL}.   Only the vector components in the plane of the orbit are shown. \vspace{0.1 in}}
%\label{fig:orbforce}
%\end{figure}

%\begin{figure}
%\includegraphics{f2.eps}
%\caption{ The surface gravity versus semi-major axis for known transiting planets.  Various theoretical mechanisms disfavor the existence of planets on the hashed sides of each labelled line.  The dashed lines are gas-giant models by \cite{2007Fa}; see text. \vspace{0.1 in}}
%\label{fig:grav}
%\end{figure}

\end{document}